\documentclass[12pt]{article}

\usepackage{graphicx}

\usepackage{amsmath,amssymb}

\usepackage[mathscr]{eucal}

\begin{document}

\title{Scanning of the $e^+e^-\rightarrow \pi^+\pi^-$
cross--section below 1 GeV by radiative events with untagged
photon}






 \author{V.A. Khoze$^1$, M.I. Konchatnij$^2$, N.P. Merenkov$^2$,
G. Pancheri$^3$, \\  L. Trentadue$^4$, O.N. Shekhovtzova$^2$}





\date{}

\maketitle

$^1$ {\small  Department of Physics and Institute for Particle
Physics Phenomenology, University of Durham, DH1 3LE, UK  and
Petersburg Nuclear Physics Institute,  Gatchina, 188350, Russia }

$^2$ {\small National Science Centre Kharkov Institute of Physics
and Technology, 61108 Akademicheskaya 1, Kharkov, Ukraine  }

$^3$ {\small INFN Laboratori Nazionali di Frascati, P.O. Box 13,
00044 Frascati, Italy  }

$^4$ {\small Dipartimento di Fisica, Universita' di Parma and
INFN, Gruppo Collegato di Parma, 43100 Parma, Italy}

\vspace{0.5cm}

\begin{abstract}

We discuss an inclusive approach to the measurement of the
$e^+e^-\rightarrow \pi^+\pi^-$ cross--section by the radiative
return method without photon tagging. The essential part of
this approach is the choice of rules for event selection which
provide rejection of events with 3 (or more) pions and decrease
the final--state radiation background. The radiative corrections
to the initial--state radiation process are computed for
DA$\Phi$NE conditions, using the quasi--real electron
approximation for both, the cross--section and the underlying
kinematics. The two cases of restricted and unrestricted
pion phase space are considered. Some numerical calculations
illustrate our analytical results.
\end{abstract} \vspace{0.2cm} PACS:  12.20.-m, 13.40.-f, 13.60.-Hb,

13.88.+e

\vspace{0.2cm}

\section{Introduction}

\hspace{0.7cm}

The recent high precision measurement of the muon anomalous
magnetic moment $(g-2)_{\mu}$ \cite{MU} has boosted interest in  renewed
theoretical calculations of this quantity
\cite{H1}, since  any difference between  the experimental value and the
theoretical evaluations based on the Standard Model (SM)   may
open a window into possible new physics \cite{HG}. While
 conclusions about posssible
discrepancy with the SM are  premature \cite{Y},  the Brookhaven based
experiment  is now planning a
new measurement with
three times better accuracy, which may create further challenges to
the theory.  Presently, there are two
main sources  of theoretical uncertainty in the calculation, namely the
impact of
the
light-by-light contribution \cite{eder,kinoshita,bijnens} and
the estimate of
the error from the hadronic vacuum polarization
contribution to $(g-2)_{\mu}$. In this paper we address the question of this
error, for which  different groups give different
results \cite{G1,G2}.

The problem of the hadronic  vacuum polarization contribution is that it
cannot be calculated analytically because perturbative QCD loses its
predictive power at low and intermediate energies, where, on the
other hand, the effect is the largest. However one can  evaluate
this hadronic term from the data on electron-positron annihilation
into hadrons by using a dispersion relation \cite{Dis}. The
necessary condition for a theoretical error matching the experimental accuracy
reached in the $(g-2)_{\mu}$ measurement, is the knowledge of the total hadronic
cross section with better than one per cent accuracy. The recent precision
measurements of the total hadronic cross-section by the CMD-2 \cite{N} and BESII
\cite{B}
collaborations were included in the new analysis of Refs.
\cite{J,J1}. While this  reduces the error in the hadronic contribution
to the shift in the running electromagnetic coupling, for the muon
(g-2) value it is
mandatory to perform new measurements of the total cross section
at energies below 1.4 GeV (in particular, in the
$e^+e^-\rightarrow \pi^+ \pi^-$- channel) with at least one per cent
accuracy. Such accurate  measurement will then be important  not just
for the
muon anomalous magnetic moment but also for testing
the effective fine structure constant.

In the last years, the idea  to use
radiative events in electron-positron collisions for scanning of
the total hadronic cross section has become quite attractive. The radiative
return approach was first discussed long ago, and   the lowest-order
cross sections for the radiative process of electron-positron
annihilation into a pair of charged fermions or scalar bosons were
calculated  \cite{BK}. This subject was subsequently studied in several
papers (see, for example, \cite{AKMT,Raz,MEMO,HGJ,mb1}), where
higher-order radiative corrections were taken into account.

Due to differences in the systematic uncertainties in the
measurement, the radiative return approach has several advantages
when compared to the conventional energy scan : for example,
luminosity and beam energy effects are accounted for only once.
For variable total hadronic energies from the $2m_{\pi}$ threshold
up to 1.02 GeV,  the ideal machine  for scanning the total cross
section, using the radiative return method, appears to be the
DA$\Phi$NE accelerator, operating at the $\Phi$ resonance,
together with the KLOE detector\cite{MEMO,Raz1,D}. DA$\Phi$NE
measurements can become quite competitive to the conventional
direct cross section scan, and,
 as   mentioned, have
certain advantages due to the systematics. The radiative return method
allows to perform precise measurements of the hadronic cross sections in
the $\rho$ resonance region. The high accuracy of scanning is
provided by the high resolution measurement of the pion 3-momenta
(and consequently the invariant mass) with the KLOE drift chamber.
Recently the first preliminary results of such measurements of
$\pi^+ \pi^-$ production cross section below 1 GeV have been
reported \cite{D}.

In our previous paper \cite{mb1} the analysis of Initial State
Radiation (ISR) effects, which provide the basis for the radiative
return
strategy, has been performed for the realistic conditions of the KLOE detector.
It was assumed that both the energy of the photon in the
calorimeter, and the
invariant mass of the $\pi^+\pi^-$ - system were measured.

Notice  that the KLOE detector allows to register photons only
outside two narrow cones along the beam directions (the so-called blind zones ).
Because of this geometrical restriction, most of the
ISR events become inaccessible for tagging and cannot be recorded
by the photon detector. This decreases the
statistics and, thus, results in lesser precision. In order to avoid
this problem and, moreover, to fully exploit the possibility of
high precision measurement of the two charged pions with the drift
chamber \cite{MEMO,Raz1}, it was proposed\footnote{Our attention
was first drawn to this idea by G.Venanzoni (see also \cite{mb1})}
to make the photon tagging redundant. The idea is to use an Inclusive
Event Selection (IES) approach, in which only the invariant mass of
the final pions is measured, and the ISR photon remains untagged.

As briefly discussed in \cite{mb1}, one of the main advantages of
the IES strategy is the rise of the corresponding cross section
caused by the $\ln(E^2/m^2)$ enhancement (here $E$ is the beam
energy and $m$ is the electron mass) due to the possibility to include
events with ISR collinear photons which, otherwise, belong to the
blind zones. As shown in \cite{GR}, the number of such untagged
photon events exceeds by about a factor  three the number of
events with the tagged photon (if the opening angle of the blind
zone equals to $10^o$).

Of course, in order to avoid uncertainties in the interpretation of
IES approach and to have the possibility to describe IES in terms
of ISR events, some additional event selection criteria should be
imposed. The corresponding additional restrictions should make the
photon tagging redundant, but at the same time should guarantee
the suppression of the main  background caused by
events from $\Phi\rightarrow\pi^+\pi^-\pi^0$ decay.

In this paper we present the  analytical calculation of
the Born cross section of ISR process
\begin{equation}\label{1}
e^-(p_1) + e^+(p_2) \rightarrow \gamma(k) + \pi^+(p_+) +\pi^-(p_-)
\end{equation}
and the QED radiative corrections (RC) to it for IES setup,
accounting for the additional kinematical constraints on the event
selection, which can be realized at KLOE. The physics motivation
for these constraints is discussed in Section 2.

The Born IES cross section is calculated in Section 3. Note that
for a chosen set of selection rules, IES cross section at the Born
level coincides with the tagged photon events cross section as
given in Ref.~\cite{AKMT}, provided that the final pion phase
space is unrestricted. But we consider also the realistic case when
the pion phase space is restricted.

In Section 4 we discuss the RC to the Born cross section caused by
the emission of real and virtual photons. At the RC level, the IES
cross section differs from the tagged photon result because of the
contribution of  double photon bremsstrahlung. The situation here
is similar to the case of radiative corrections  in  DIS with
detected lepton (the analogue of the tagged photon events) or
hadrons (analogous to the IES). In the latter case, the radiative
corrections factorize while in the former they include by
necessity some integrals over hadronic cross section that have to
be extracted from experimental data. This fact certainly makes the
IES approach more advantageous. In Section 5 the cancellation of
the infrared and collinear parameters,  used in the calculations
of radiative corrections,
 is
demonstrated, and
the expression for the total photonic contribution to the radiative corrections
 is
given.
In Section 6 we discuss also possible contribution of the
$e^+e^-$--pair production into the IES cross section if the
$e^+e^- \pi^+\pi^-$ final state is not rejected by the analysis
procedure. Our Conclusion contains a brief summary and  the
discussion of the background processes which may contribute into
the IES cross section.


\section{IES selection rules}

\hspace{0.7cm}

As mentioned in the Introduction, the main condition of the IES
approach is the precise measurement of the di--pion invariant mass
in process (1). In addition,  restrictions must be imposed in order to
select final states with only
$\pi^+\pi^- +n \gamma$ , excluding $\pi^+\pi^-
\pi^0$. Finally,  we have to add some
constraints in order to  reduce contributions from final state radiation
  (FSR). As an added bonus, such contraints also simplify  the
 theoretical calculation of the radiative corrections.

Rejection of the 3--pion final state in process (1) can be done
selecting events with an appropriately small  difference between the lost
(undetected) energy and the modulus of the lost 3--momentum in
process (1).
In terms of the measured pion 3--momenta, this restriction
reads \cite{MEMO, mb1, Raz1}
\begin{equation}\label{2}
2E-E_+-E_- -|{\bf P}_{\Phi}-{\bf p_+ - p_-}| < \eta E\ , \ \ \eta \ll 1,
\end{equation}
where $E$ is the beam energy, $E_{\pm}= \sqrt{{\bf
p}_{\pm}^2-m_{\pi}^2}$ is the energy of $\pi^{\pm}$, and $m_{\pi}$
is the pion mass. ${\bf P}_{\Phi}$ is the total initial--state
3--momentum which, at DA$\Phi$NE, is non-zero, due to a small
acollinearity in the beams, $|{\bf P}_{\Phi}| = 12.5 MeV.$  If the
chosen parameter $\eta$ is small enough ($\leq (m_\pi/E)^2)$,
constraint (2) allows to avoid the undetected $\pi^0$ and to
retain only the undetected $n \gamma$ system. Inequality (2) can
be rewritten in terms of the total energy $\Omega$ and modulus of
the total 3--momentum $|{\bf K}|$ of all photons in the reaction
$e^+ + e^- \rightarrow \pi^+ + \pi^- +n\gamma$ as
\begin{equation}\label{3} \Omega- |{\bf K}| < \eta E.
\end{equation} The optimal value $\eta = 0.02$ decreases also the FSR
background \cite{MEMO}.

The next constraint
selects such events, where at  $n=1$ the undetected
photon is collinear with the emitting electron (or positron). The
collinear events considered here are those in which the photon
belongs to a narrow cone with  opening angle $2\theta_0 \
(\theta_0 \ll 1)$ along the electron beam direction (the blind zone for the
KLOE detector). This
constraint reads
\begin{equation}\label{4}
{\bf Kp_1 > |K|}Ec_0, \ \ c_0 =\cos{\theta_0}\ ,
\end{equation}
where ${\bf p_1}$ is the 3--momentum of the electron and
$\theta_0$ can be chosen to be  $5^o\div 6^o$. Due to this
constraint, the collinear photon radiated by the initial electron
contributes to the observed IES cross section and induces a
$\ln(E^2\theta_0^2/m^2)$ enhancement, which,  at
DA$\Phi$NE, makes  the IES cross section  a few times larger
than in  the tagged photon case. Moreover, the collinear
constraint provides the possibility to apply the well known
quasi--real electron (QRE) method \cite{BFK} to calculate radiative
corrections. Even
at the Born level, the difference between the exact result and the
corresponding QRE approximation is negligible (see for details
Section 3). Selection rules (2), (3) and (4) imply precise
measurements of the pion 3--momentum that can be provided by the KLOE
drift chamber.

Note also that in the Born approximation ($n=1$) inequality (3) is
always satisfied, and, therefore, the  selection rules (3) and (4)
imply non--trivial consequences only through the contribution to the
radiative corrections from
two hard photon emission.

Because of the existence of the blind zones, the KLOE detector
cannot provide the detection of the final $\pi^+$ and $\pi^-$
inside the full  phase space, picking  out events with pion
polar angles in the region
\begin{equation}\label{5}
\theta_m < \theta_{\pm} < \pi -\theta_m\ .
\end{equation}

In principle, $\theta_m$ can be taken to be  about $10^o$, but, as
shown by  the Monte Carlo calculations \cite{Raz,MEMO}, the choice
of $\theta_m$ influences also the value of the FRS background. The
optimal value of $\theta_m$ for DA$\Phi$NE conditions is $20^o.$
\cite{MEMO}

Usually, the restricted pion phase space can be taken into account by
introducing  an  acceptance factor $A(\theta_m).$ The
calculation of this factor is very simple for a nonradiative
process, but for the ISR process, and the RC to its cross section,
it is non--trivial. The analytical form of $A(\theta_m)$ can be
derived in the framework of QRE approximation. To illustrate the
problem, in the following
 we perform calculations for both unrestricted and restricted
 pion phase space.


\section{Born approximation}

\hspace{0.7cm}

To lowest order in $\alpha,$ the differential cross section of
process (1) can be written in terms of the leptonic $L_{\mu\nu}$
and hadronic $H_{\mu\nu}$ tensors as (see Ref.~\cite{BK})
\begin{equation}\label{6}
d\sigma^B = \frac{8\pi^2\alpha^2}{sq^4}L^{\gamma}_{\mu\nu}(p_1,p_2,k)
H_{\mu\nu}\frac{\alpha}{4\pi^2}\frac{d^3k}{\omega}\frac{d^3p_+d^3p_-}
{16\pi^2E_+E_-}\delta(q-p_+-p_-)\ ,
\end{equation}
where $\omega$ is the energy of photon, and the hadronic tensor is
expressed via the pion electromagnetic form factor $F_{\pi}(q^2)$
as follows $$H_{\mu\nu} = -4|F_{\pi}(q^2)|^2\tilde p_{-\mu}\tilde
p_{-\nu}\ , \ \ \tilde p_{-\mu} = p_{-\mu}-\frac{1}{2}q_{\mu}\ , \
\ q=p_1+p_2-k =p_++p_- \ . $$

The pion electromagnetic form factor defines the total cross section $\sigma
(q^2)$ of the process $e^++e^-\rightarrow \pi^++\pi^-$, by means of relation
\begin{equation}\label{7}
|F_{\pi}(q^2)|^2=\frac{3q^2\sigma(q^2)}{\pi\alpha^2\zeta}, \ \ \zeta=\Bigl(
1-\frac{4m_{\pi}^2}{q^2}\Bigr)^{\frac{3}{2}} \ .
\end{equation}
In the case of unrestricted (full) pion phase space, the  integration of the
hadronic tensor can be performed in invariant form \cite{BK}
\begin{equation}\label{8}
\frac{1}{16\pi^2}\int
H_{\mu\nu}\frac{d^3p_+d^3p_-}{E_+E_-}\delta(q-p_+-p_-)
=\frac{q^4}{8\pi^2\alpha^2}\sigma(q^2)\tilde g_{\mu\nu}, \ \
\tilde g_{\mu\nu}=g_{\mu\nu}-\frac{q_{\mu}q_{\nu}}{q^2}\ ,
\end{equation}
whereas for the  restricted case we have first to contract the
tensors in Eq.~(6) and then to integrate the result over the pion
phase space.

The leptonic tensor on the right--hand side of Eq.~(6) for the
case of collinear ISR along the electron beam direction is well
known \cite{BK,KMF}
\begin{equation}\label{9}
L^{\gamma}_{\mu\nu}(p_1,p_2,k)=\Bigl[\frac{(q^2-t_1)^2+(q^2-t_2)^2}
{t_1t_2}-\frac{2m^2q^2}{t_1^2}\Bigr]\tilde g_{\mu\nu}+\frac{4q^2}{t_1t_2}
\tilde p_{1\mu}\tilde p_{1\nu} + \Bigl(\frac{4q^2}{t_1t_2}-\frac{8m^2}
{t_1^2}\Bigr)\tilde p_{2\mu}\tilde p_{2\nu},
\end{equation}
where $$\tilde p_{1,2\mu} = p_{1,2\mu}
-\frac{p_{1,2}q}{q^2}q_{\mu}\ , \ t_1=-2kp_1, \ t_2=-2kp_2, \
s=2p_1p_2, \ q^2=s+t_1+t_2, $$ and we neglect terms of the order
$|m^2/t_2|$ which are always below the required
accuracy.\footnote{Note that formula (9) corresponds to the
radiation along the electron beam direction (see also inequality
(4)). To account for the  radiation along the positron we have to
multiply all results for the cross sections by a factor of 2 (see
the end of Section 6).} The contraction of the leptonic tensor
with $\tilde g_{\mu\nu}$ that is necessary to use in the case of
unrestricted pion phase space reads
\begin{equation}\label{10}
L^{\gamma}_{\mu\nu}(p_1,p_2,k)\tilde g_{\mu\nu}=2F\ , \ \
F=\frac{(q^2-t_1)^2+(q^2-t_2)^2}{t_1t_2}-\frac{2q^2m^2}{t_1^2}\ .
\end{equation}

In the case of restricted phase space we have to use the following
relation
\begin{equation}\label{11}
L^{\gamma}_{\mu\nu}(p_1,p_2,k)H_{\mu\nu}=4q^2|F_{\pi}(q^2)|^2 R \ ,
\end{equation}
$$R=-\frac{m_{\pi}^2}{q^2}F
+\frac{q^2(\chi_1+\chi_2)-\chi_1^2-\chi_2^2}
{t_1t_2}-\frac{\chi_1}{t_1}-\frac{\chi_2}{t_2}+\frac{2m^2\chi_2}{t_1^2}
\Bigl(\frac{\chi_2}{q^2}-1\Bigr)\ , \ \ \chi_{1,2}=2p_{1,2}p_-\
.$$ Therefore, the differential cross section of the process (1)
in the Born approximation can be written as
\begin{equation}\label{12}
d\sigma^B_F=\sigma(q^2)\frac{\alpha}{2\pi^2}F\frac{d^3k}{\omega}\ ,
\end{equation}
\begin{equation}\label{13}
d\sigma^B_R=
\frac{3\sigma(q^2)R}{s\zeta}\frac{\alpha}{4\pi^2}\frac{d^3k}{\omega}\frac
{|{\bf
p}_-|dE_-dc_-}{E_+}\frac{d\varphi_-}{2\pi}\delta(2E-\omega-E_+-E_-)\
, \end{equation} where $c_-=\cos{\theta_-}$ and $\theta_-, \
\varphi_-$ are polar and azimuthal angles of the negative pion
respectively (we take Z axis along the ${\bf p_1}$ direction).

It seems at first sight that one can perform the trivial integration
with respect to the azimuthal angle $\varphi_-$ on the right--hand
side of Eq.~(13) because the quantity $R$ does not contain any
$\varphi_-$--depending term. But in the general case  the pion
energies $E_+$ and $E_-$ depend on $\varphi_-.$ Moreover, the
upper limit of integration over $c_-$ for the ISR events with
collinear photon along the electron beam direction is smaller than
$c_m = \cos{\theta_m}$ and depends on $\varphi_-$ as well, i.e.
\begin{equation}\label{14a}
-c_m <c_- <c_{max}\ , \ \ \ c_{max}<c_m\ ,
\end{equation}
where $c_{max}$ must be determined from the 3--momentum
conservation, provided that $c_+ =\cos{\theta_+} = -c_m.$
Therefore, in the case of the restricted pion phase space, it is
necessary  to  first integrate over $c_-$ and then over
$\varphi_-$.

Let us now perform the integration with respect to the photon
angular phase space on the right--hand side of Eq.~(12). It is
convenient to choose $X$ axis along the ${\bf P}_{\Phi}$
direction. In the laboratory frame
\begin{equation}\label{15}
p_1 = (E, 0,0,|{\bf p_1}|)\ , \ \ p_2 = (E,|{\bf P}_{\Phi}|,0,-p_{2z})\ , \
\ p_{2z} = E\bigl(1-\frac{{\bf P}_{\Phi}^2}{2E^2}\bigr)
\end{equation}
and
$$t_1=-2\omega(E-|{\bf p_1}|\cos{\theta})\ , \ \
t_2=-2\omega E\bigl[1+\bigl(1-\frac{{\bf
P}_{\Phi}^2}{2E^2}\big)\cos{\theta}\bigr]+
2\omega|{\bf P}_{\Phi}|\sin{\theta}\cos{\phi}\bigr]\ , $$
\begin{equation}\label{16}
s=4E^2-{\bf P}_{\Phi}^2, \ \ q^2= 4E^2-4E\omega -{\bf P}_{\Phi}^2\bigl(1-
\frac{\omega}{E}\cos{\theta}\bigr)+2\omega|{\bf P}_{\Phi}|\sin{\theta}
\cos{\phi}\ ,
\end{equation}
where $\theta$ and $\phi$ are polar and azimuthal angles of the
photon radiated in the initial state. Keeping terms of the order
${\bf P}_{\Phi}^2/E^2$ and $\theta_0^2,$ we can use the following
list of integrals
$$\int\frac{d\,O}{-t_2}=\frac{\pi\theta_0^2}{4\omega_0E}\ , \
\int\frac{d\,O}{-t_1} = \frac{\pi}{\omega_0 E}\Bigl[\Bigl(1-
\frac{{\bf P}_{\Phi}^2}{4E^2}\Bigr)L_0
-\frac{\theta_0^2}{12}\Bigr]\ , \ \int\frac{d\,O m^2}{t_1^2}=
\frac{\pi}{\omega_0^2}\Bigl(1- \frac{{\bf
P}_{\Phi}^2}{2E^2}\Bigr)\ , $$
\begin{equation}\label{17}
\int\frac{d\,O t_2}{t_1}=4\pi\Bigl[\Bigl(1-
\frac{{\bf P}_{\Phi}^2}{4E^2}\Bigr)L_0 -\frac{\theta_0^2}{3}\Bigr]\ , \
\ \ \int\frac{d\,O}{t_1t_2}=\frac{\pi}{4\omega_0^2E^2}\Bigl[\Bigl(1-
\frac{{\bf P}_{\Phi}^2}{4E^2}\Bigr)L_0 +\frac{\theta_0^2}{6}\Bigr]\ ,
\end{equation}
$$\omega_0=\frac{4E^2-q^2-{\bf P}_{\Phi}^2}{4E}\ , \ \ L_0 =
\ln{\frac {E^2\theta_0^2}{m^2}}\ , \ \ d\,O=d\cos{\theta}d\phi\ .
$$ When evaluating these integrals we systematically neglected
small terms of order ${\bf P}_{\Phi}^2\theta_0^2/E^2.$ Within such
approximation we can use the substitution
\begin{equation}\label{18}
|{\bf P}_{\Phi}|\sin{\theta}\cos{\phi} \rightarrow 0, \ \ \ {\bf
P}_{\Phi}^2\cos{\theta} \rightarrow {\bf P}_{\Phi}^2\ ,
\end{equation}
in the expressions for the invariants $t_1$ and $t_2$ in Eqs.~(17). With
the same accuracy we can write
\begin{equation}\label{19}
\omega d\omega=\frac{\omega_0dq^2}{4E}\Bigl(1+\frac{{\bf
P}_{\Phi}^2}{2E^2}\Bigr)\ .
\end{equation}

Combining (12), (17) and (19), we arrive at the distribution over
the pion squared invariant mass $q^2$, for unrestricted pion phase
space
\begin{equation}\label{20, Born}
\frac{d\sigma^{^B}_F}{dq^2}=\frac{\sigma(q^2)}{4E^2}\frac{\alpha}{2\pi}\Bigl[
\Bigl(\frac{q^4}{8\omega_0E}+\frac{q^2}{2E^2}+\frac{\omega_0}{E}\Bigr)\Bigl(
1+\frac{{\bf P}_{\Phi}^2}{4E^2}\Bigr)L_0
-\frac{q^2}{2\omega_0E} +
\frac{\theta_0^2}{6}
\Bigl(\frac{q^4}{8\omega_0E}+\frac{q^2}{2E^2}-\frac{2\omega_0}{E}\Bigr)
\Bigr]\ .
\end{equation}
To guarantee only one per cent accuracy one can neglect terms
proportional to ${\bf P}_{\Phi}^2/4E^2$ and $\theta_0^2/6$ in (20)
because their contribution into the IES cross section is of the
relative order $10^{-4}.$ Such procedure leads to the well known
result corresponding to the QRE approximation \cite{BFK}
\begin{equation}\label{21}
\frac{d\sigma^{^B}_F}{dq^2}=\frac{\sigma(q^2)}{4E^2}\frac{\alpha}{2\pi}P(z,L_0)\
, \ P(z,L_0) = \frac{1+z^2}{1-z}L_0-\frac{2z}{1-z}\ , \
z=\frac{q^2}{4E^2}\ .
\end{equation}

Thus, the QRE approximation appears to be sufficient for a description
of the IES cross section and  corresponds to a one per cent
precision even at the Born level.

Consider now the case of the restricted pion phase space. First, the
$\delta$--function  on the right--hand side of Eq.~(13) has to be
used to perform the integration with respect to $E_-.$ Then within
chosen accuracy we obtain
\begin{equation}\label{22}
\frac{|{\bf p}_-|d\,E_-}{E_+}\delta(2E-\omega-E_+-E_-)=\frac{|{\bf
p}_-|^2}{|{\bf p}_-|(2E-\omega)+E_-
(\omega c_{\gamma_-}-|{\bf
P}_{\Phi}|s_-\cos{\varphi_-})}\ ,
\end{equation}
where
$$c_{\gamma_-}=\cos{\theta}c_-+\sin{\theta}s_-\cos{(\phi-\varphi_-)}\
, \ \ s_- = \sin{\theta_-}\ , \ \
\omega=\omega_0\Bigl(1+\frac{{\bf P} _{\Phi}^2}{4E^2}\Bigr) \ . $$
Now let us express $E_-$ in terms of the photon energy $\omega$
and angles of the photon and negative pion, using the energy--momentum
conservation. The result can be written in the following form
\begin{equation}\label{23}
E_- = \frac{AB-C\sqrt{B^2-4m_{\pi}^2(A^2-C^2)}}{2(A^2-C^2)}\ , \ \
A=2E-\omega, \ \ B=4E(E-\omega)-|{\bf P}_{\Phi}|^2\ ,
\end{equation}
$$C =\omega c_{\gamma_-}-|{\bf P}_{\Phi}|s_-\cos{\varphi_-}\ .$$

To find the energy of the positive pion it is necessary to use the
relation $2E-\omega = E_++E_-.$ Having expressions for the pion
energies, we can apply conservation of the Z--component of
3--momentum at $c_+ = -c_m$
\begin{equation}\label{24}
-|{\bf p}_+|c_m + |{\bf p}_-|c_- +\omega\cos{\theta} =0
\end{equation}
to derive the upper limit $c_{max}$ of the variable $c_-.$  In
Eq.~(24) both $|{\bf p}_-|$ and $|{\bf p}_+|$ are functions of
$\omega, \ c_-, \ \cos{\theta}$ and $\cos{(\phi-\varphi_-)}.$
Therefore, to reach the same accuracy as in Eq.~(20) for the case
of the restricted pion phase space, the differential distribution
over the pion invariant mass squared has to be taken in the form
\begin{equation}\label{25}
\frac{d\sigma^{^B}_R}{d\,q^2}= \frac{12\sigma(q^2)}{4E^2\xi}\frac{\alpha}
{2\pi}\frac{\omega_0}{E}\Bigl(1+\frac{{\bf P}_{\Phi}^2}{2E^2}\Bigr)
\int\limits_{c_0}^1d\cos{\theta}\int\limits_0^{2\pi}\frac{d\phi}{2\pi}
\int\limits_0^{2\pi}\frac{d\varphi_-}{2\pi}\int\limits_{-c_m}^{c_{max}}
d\,c_-\frac{R|{\bf p}_-|^2}{A|{\bf p}_-|+CE_-}\ ,
\end{equation}
where $c_0 =\cos{\theta_0}$ and the quantities $A, \ C$ are defined in
(23). The analytical integration on the right--hand side of
Eq.~(25) is not available  and  the  task of integration can be
left for numerical calculation.

The result is very much simplified if one neglects terms of the
order $|{\bf P}_{\Phi}^2|/E^2$ and use the QRE approximation,
assuming $\cos{\theta} =1$ in  expressions (22), (23) and
(24). This leads to the IES cross section
\begin{equation}\label{26}
\frac{d\sigma^{^B}_R}{d\,q^2}=\frac{\sigma(q^2)}{4E^2}\frac{\alpha}{2\pi}
P(z,L_0)A(z,c_m)\ ,
\end{equation}
where $z$ is defined in (21) and
\begin{equation}\label{27}
A(z,c_m)
=\frac{12}{\zeta}\int\limits_{-c_m}^{c_{max}}d\,c_-U
\frac{z[(1+z)K-(1-z)c_-]^2}{K[(1+z)^2-(1-z)^2c_-^2]^2}\ , \ U=\frac
{\chi_1}{4E^2}-\frac{\chi_1^2}{16E^4}-\frac{m_{\pi}^2}{4zE^2}
\end{equation}
represents the  acceptance factor (see the end of the previous
Section) as it follows from the comparison between the IES cross
sections (21) and (26). To write down $A(z,c_m)$ the following
notations were used:
$$K=\sqrt{1-\frac{\delta^2}{z^2}[(1+z)^2-(1-z)^2c_-^2]}\ , \ \
\frac
{\chi_1}{4E^2}=\frac{z[1+z-2Kc_-+(1-z)c_-^2]}{(1+z)^2-(1-z)^2c_-^2}\
, \ \delta^2 =\frac{m_{\pi}^2}{4E^2} \ , $$
\begin{equation}\label{28}
c_{max}(z,c_m) =\frac{(1+z)g}{\sqrt{(z-(1-z)g)^2-(1+z)^2\delta^2}}\ , \
g=\frac{zc_m[(1+z)K(c_m)+(1-z)c_m]}{(1+z)^2-(1-z)^2c_m^2}-\frac{1-z}{2}\ .
\end{equation}

At fixed values of $c_m$, the quantity $c_{max}$ depends on the
squared pion invariant mass $q^2=4E^2z.$ If $z$ is small enough
(this situation corresponds to the radiation of a very hard
collinear photon with the energy fraction $(1-z)$ by the electron)
$c_{max}$, as formally defined by (28), can approach $-c_m$ and
became even smaller. Because selection rule (5) forbids any values
of $c_{max}$ smaller than $c_m,$ it is necessary to substitute the
upper limit of integration on the right--hand side of Eq.~(27) by
$max[c_{max}, \ -c_m].$ If not, the formal calculation in (27)
leads to negative values for the acceptance factor at small $z,$
as one can see from from Fig.~1 for $c_m=\cos{ 20^o}$ and
$z\approx 0.1$, while really it equals to zero at such
$z$--values. Note that in the framework of the QRE approximation
one can use also
 energy conservation for events with $\theta_+ = \pi
-\theta_m$ $$E(1+z)=E_-(c_{max})+E_+(c_+=-c_m)$$ in order to
obtain the analytical form of $c_{max}$ on the Born level.

If $c_m=1$,  3--momentum conservation in  process (1)
requires $c_{max}(z,1)=1$ as well. In this case
$$K(c_m)=\sqrt{1-\frac{4\delta^2}{z}}\ , \ \
g=\frac{(1+z)K(c_m)-1+z}{4}, $$ and it is easy to see that
$c_{max}(z,c_m)$, as given by (28), satisfies this requirement.

In principle, the acceptance factor $A(z,c_m)$ may be computed
analytically by means of Euler's substitution on the right--hand
side of Eq.(27) $$c_-=\frac{t^2-a^2}{2t}, \ \
K=\frac{\delta(1-z)(t^2+a^2)}{2zt}, \ \ dc_-=
\frac{(t^2+a^2)dt}{2t^2}, \ \
a^2=\frac{z^2-\delta^2(1+z)^2}{\delta^2(1-z)^2}. $$ By definition,
$A(z,1)=1$ and for $c_m<1$ always $A(z,m)<1.$

We leave the task of the analytical integration of the acceptance
factor in general case aside and first only note that the limit
$\delta\rightarrow 0$ (that is, of course, not the case for
DA$\Phi$NE) can be used to control our calculations. In this
limiting case
$$A(z,K=1,1)=12z^2\int\limits_{-1}^1\frac{dx(1-x^2)}{[1+z+(1-z)x]^4}
=1\ .$$

The acceptance factor as a function of the pion
squared invariant mass is shown in Fig.~1 for $\theta_m =10^o$ and
$\theta_m=20^o.$ We see that the acceptance factor  $A(z,c_m)$ is
close to unity in a wide z-range, but decreases very rapidly with
the  pion invariant mass.

\begin{figure}[h]
\includegraphics[width=0.47\textwidth]{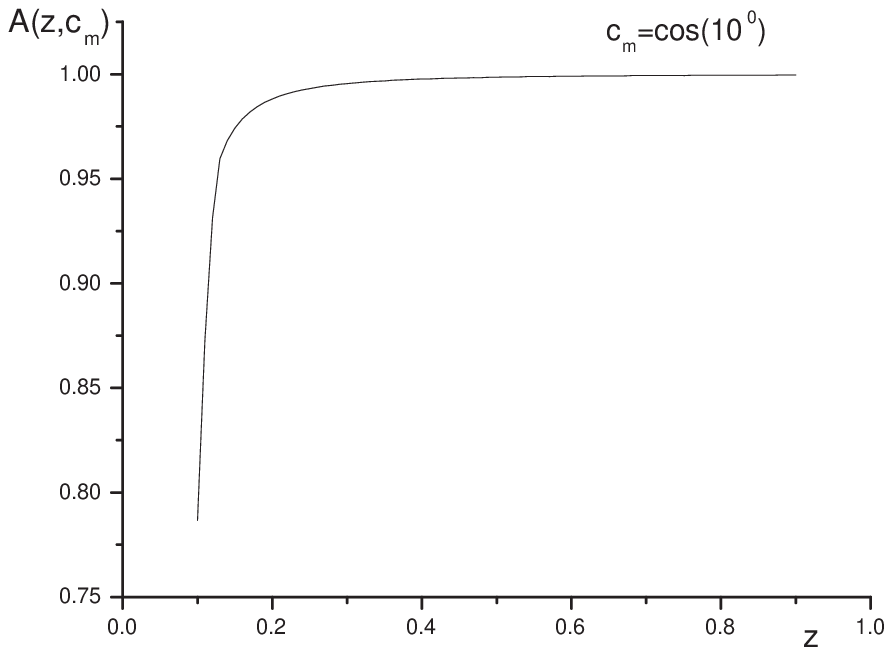}
\includegraphics[width=0.47\textwidth]{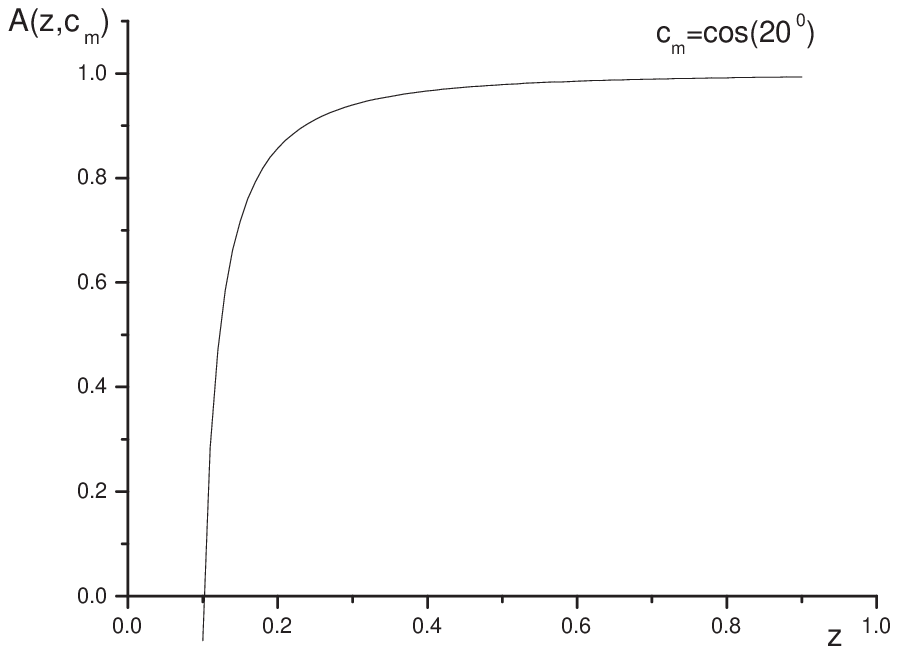}
 \parbox[t]{1\textwidth}{\caption{Acceptance factor defined by Eq.~(27)
 for different values of $c_m$ }\label{fig1}}
\end{figure}

Finally, let us note that two representations (21) and (26) for
the IES cross section can be obtained also by inserting the QRE
form of the leptonic tensor
$$\frac{\alpha}{4\pi^2}\frac{d^3k}{\omega}L^{\gamma}_{\mu\nu}(p_1,p_2,k)
\rightarrow
\frac{\alpha}{2\pi}P(z,L_0)\frac{dz}{z}q^2\bigl(\frac{1}{2} \tilde
g_{\mu\nu}+\frac{2z^2}{q^2}\tilde p_{1\mu}\tilde p_{1\nu}\bigr) $$
into Eq.~(6).

\section{Radiative corrections}

\hspace{0.7cm}

If events with  $e^+e^-\pi^+\pi^-$ final state are rejected,
only photonic RC have to be taken into account. These corrections
include contributions due to virtual and real soft and hard photon
emission. To calculate them we use the QRE approximation from the
very beginning.
\subsection{Soft and virtual corrections}
The soft and virtual corrections are the same for
both, unrestricted and restricted pion phase space, and the corresponding
contribution can be found by the simple substitution
\begin{equation}\label{29}
\frac{\alpha}{2\pi}P(z,L_0)\rightarrow\Bigl(\frac{\alpha}{2\pi}\Bigr)^2
C^{S+V}\ , \ \ C^{S+V} = \rho P(z,L_0) +D(L_s,L_0,z) +N(z)
\end{equation}
in the right--hand sides of Eqs.~(21) and (26). As a result, we
have
\begin{equation}\label{30}
\frac{d\sigma^{^{S+V}}_F}{dq^2}=\frac{\sigma(q^2)}{4E^2}
\Bigl(\frac{\alpha}{2\pi}\Bigr)^2C^{S+V}\ , \ \
\frac{d\sigma^{^{S+V}}_R}{dq^2}=\frac{d\sigma^{^{S+V}}_F}{dq^2}A(z,c_m)\ .
\end{equation}

All the logarithmically  enhanced  contributions to $C^{S+V}$ are
contained in the first two terms in Eq.(29) and were first found in
Ref.~\cite{AKMT}.\footnote{There is a misprint in the expression
for $\rho$ given in Ref.~\cite{AKMT}. The correct expression includes an
additional $\pi^2$ term.} The third term in (29) describes the
non--logarithmic contributions \cite{UZ}. Finally, we obtain
\begin{equation}\label{31}
\rho=4(L_s-1)\ln{\Delta}+3(L_s+\ln{z})+\frac{2\pi^2}{3}-\frac{9}{2}\
, L_s=\ln{\frac{4E^2}{m^2}}\ ,
\end{equation}
$$ D(L_s,L_0,z)=
\frac{1+z^2}{1-z}L_0\bigl[(L_0-2L_s-\ln{z})\ln{z}+\frac{\pi^2}{3}
-2Li_2(z)\bigr] +\frac{1+2z-z^2}{2(1-z)}L_0
+\frac{4z\ln{z}}{1-z}L_s\ ,$$ $$
N(z)=-\frac{1}{1-z}-\frac{8z\ln{z}}{1-z}+2z\bigl[\ln^2(1-z)+\frac{\ln^2{z}}
{1-z}\bigr] +\frac{\pi^2}{6}\bigl(4z+6-\frac{5}{1-z}\bigr) +
\bigl(4z-6+\frac{5}{1-z}\bigr)Li_2(z)\ , $$ where $\Delta E \
(\Delta \ll 1)$ is the maximum energy of a soft photon, and
$$Li_2(x) = -\int\limits_0^x\frac{dy}{y}\ln(1-y)\ .$$

\subsection{Two hard photon emission along the electron and positron  direction}
Concerning the contribution from additional hard
photon emission, we divide it into three pieces. The first one is
responsible for the radiation of an additional photon with  energy
$\omega_2$ along the positron beam direction (provided that a
collinear photon with  energy $\omega_1$ is emitted along the
electron beam direction). To calculate it we introduce the angular
auxiliary parameter $\theta_0'\ll 1 $ and use the QRE
approximation to describe the radiation of both photons. The
calculations are the same for restricted and unrestricted pion phase
space and do not affect the acceptance factor (as  in the case of
virtual and soft corrections). Using the subscript F to indicate the
cross-section for the unrestricted  phase space, the result reads
\begin{equation}\label{32}
\frac{d\sigma^{H}_{1F}}{dq^2}=\frac{\sigma(q^2)}{4E^2}
\int_{y_0}^{1-\Delta} \Bigl(\frac{\alpha}{2\pi}\Bigr)^2
P(x,L_0)P(y,L_0')\frac{dy}{y}\ ,
\ L_0'=
\ln{\frac{E^2{\theta'}_0^2}{m^2}}, \ x=1-\frac{\omega_1}{E}, \ y=1-\frac
{\omega_2}{E},
\end{equation}
where $q^2 = 4E^2xy$ is fixed. The maximum value of the photon
energy $\omega_2$ can be obtained from restriction  (3), taking
into account that, for the events under consideration, the QRE
kinematics $(\cos{\theta_1}=1, \ \cos{\theta_2}=-1, \
\theta_{1,2}$ are polar angles of photons) gives
\begin{equation}\label{33}
\Omega=\omega_1+\omega_2, \ \ |{\bf K}|=\omega_1-\omega_2, \
\rightarrow \omega_2<\frac{\eta E}{2}, \ \ y_0 =1-\frac{\eta}{2}.
\end{equation} Because of the smallness of the parameter $\eta$, only
terms singular at $y=1$  contribute to the integral on the
right--hand side of Eq.~(32). Moreover, we can substitute $x$ with
$z$, neglecting terms of order $\eta$ as compared with unity
(because $q^2 = 4E^2z =4E^2xy)$, and perform an elementary
integration over $y$, obtaining
\begin{equation}\label{34}
\frac{d\sigma^{^H}_{1F}}{dq^2}=\frac{\sigma(q^2)}{4E^2}\Bigl(\frac{\alpha}
{2\pi}\Bigr)^2P(z,L_0)2(L_0'-1)\ln{\frac{\eta}{2\Delta}}\ , \ \
\frac{d\sigma^{^H}_{1R}}{dq^2}=\frac{d\sigma^{^H}_{1F}}{dq^2}A(z,c_m).
\end{equation}

\subsection{Two hard photons emitted along the electron directions}

The second  contribution to the radiative corrections caused by an
additional hard photon emission corresponds to the radiation of
two hard collinear photons (each of them with  energy larger
than $\Delta E$) by the electron, provided  both are emitted
within the narrow cone of opening angle $2\theta_0$ along
the electron beam direction. This contribution
also does not affect the acceptance factor, and the result can be
written with the same accuracy as (30) and (32)
\begin{equation}\label{35}
\frac{d\sigma^{^H}_{2F}}{dq^2}=\frac{\sigma(q^2)}{4E^2}
\Bigl(\frac{\alpha}{2\pi}\Bigr)^2C^{^H}\ , \ \
\frac{d\sigma^{^H}_{2R}}{dq^2}= \frac{d\sigma^{^H}_{2F}}{dq^2} A(z,c_m)\ ,
\end{equation}
$$C^{^H} =B_1(z,\Delta)L_0^2+B_2(z,\Delta)L_0+B_3(z,\Delta)\ .$$
Functions $B_1(z,\Delta)$ and $B_2(z,\Delta)$ were first
calculated in Ref.~\cite{M} $$B_1(z,\Delta) =
\frac{1}{2}P_{2\theta}(z) +\frac{1+z^2}{1-z}\bigl(\ln z
-\frac{3}{2}-2\ln\Delta\bigr), $$
\begin{equation}\label{36}
B_2(z,\Delta)=3(1-z)+\frac{(3+z^2)\ln^2z}{2(1-z)}-\frac{2(1+z)^2}{1-z}\ln
\frac{1-z}{\Delta}\ ,
\end{equation}
where $P_{2\theta}(z)$ is the $\theta$--term of the second order
electron structure function (see, for example, \cite{JSW}) $$
P_{2\theta}(z)=2\Bigl[\frac{1+z^2}{1-z}\Bigl(2\ln(1-z)-\ln{z}+\frac{3}{2}
\Bigr)+\frac{1}{2}(1+z)\ln{z}-1+z\Bigr]\ . $$ Function
$B_3(z,\Delta)$ was calculated in Ref.~\cite{UZ} and  reads
\begin{equation}\label{37}
B_3(z,\Delta)=\frac{4z}{1-z}\ln\frac{1-z}{\Delta}-\frac{4z}{3(1-z)}-\frac
{\pi^2}{6}\frac{4z}{3}+\Bigl(-\frac
{17}{3}+\frac{28}{3(1-z)}-\frac{8}{3(1-z)^2}\Bigr)\ln{z}+
\end{equation}
$$+\frac{3-24z+54z^2-48z^3+7z^4}{6(1-z)^3}\ln^2z
+\bigl(1-\frac{7z}{3}\bigr)Li_2(1-z)+J\ , $$
$$J=\int\limits_0^{1-z}\Bigl[\frac{z^2+(1-x)^4}{x\lambda(1-x)^2}\int
\limits_0^1\frac{dt}{t}L(t,x,z)
+\frac{z+x}{2(1-x)}L_1(x,z)+\frac{x\lambda-3z}{2(1-x)^2}L_2(x,z)\Bigr]dx,
$$ where the logarithmic functions $L,\ L_1$ and $L_2$ are defined
as follows
$$L(t,x,z)=\ln{\frac{(1-x)\sqrt{F(t,x,z)}+tx(z-x\lambda)+\lambda(1-x)^2}
{2\bigl(1+\frac{\lambda t}{xz}\bigr)\lambda(1-x)^2}}\ , \ \
\lambda =1-x-z ,$$
$$L_1(x,z)=\ln{\frac{(x+z))\sqrt{F(1,x,z)}+\lambda(z-x\lambda)+x(x+z)^2}
{2z\lambda}}\ , $$
$$L_2(x,z)=\ln{\frac{(1-x)\sqrt{F(1,x,z)}+x(z-x\lambda)+\lambda(1-x)^2}
{2\lambda(1-x)^2}}\ , $$ $$F(t,x,z)=
\lambda^2(1-x)^2+2tx\lambda(z-x\lambda)+t^2x^2(x+z)^2\ .  $$

The integral over $x$ in the expression for $J$ diverges when
$x\rightarrow 0$ and $x\rightarrow 1-z.$ But the structure of the
integral is such that these divergences compensate each other, and
this can be seen by taking into account that:

\centerline{\it i) \ integral
over $t$ converges,}

 $$ii) \ \lim\limits_{x\to 0} \
\frac{z^2+(1-x)^4}{x\lambda(1-x)^2}L(t,x,z)=
-\frac{1+z^2}{x(1-z)}\ln{\frac{t(1-z)}{xz}}\ , $$
\begin{equation}\label{38}
iii)\ \lim\limits_{x\to 1-z} \
\frac{z^2+(1-x)^4}{x\lambda(1-x)^2}L(t,x,z)=
\frac{1+z^2}{\lambda(1-z)}\ln{\frac{t(1-z)}{(1-x-z)z}}\ .
\end{equation}

The quantity $J$ can then be computed numerically with
appropriate precision. Note also that the integral over $t$ on the
right--hand side of Eq.~(37) can be computed analytically (see
also \cite{A}). i.e.
 $$\int\limits_0^1\frac{dt}{t}L(t,x,z) =
\frac{1}{2}L_2^2(x,z)+Li_2\Bigl(\frac{2xz}{h}\Bigr)+Li_2\Bigl(-\frac{2x^2
\lambda}{h}\Bigr) +Li_2\Bigl(-\frac{\lambda}{xz}\Bigr)\ , $$
$$h=\lambda(1-x)^2+x(z-x\lambda)+(1-x)\sqrt{F(1,x,z)}\ .$$

\subsection{Emission of one hard collinear and one hard wide angle
photon with unrestricted pion phase space}
The third, least trivial,  contribution into the RC caused by
two hard photon emission is connected with events when one
photon with energy $\omega_1$ is collinear and the other (with
energy $\omega_2$) is radiated at  angles between
$\pi-\theta_0'$ and $\theta_0.$ For such events,  constraints
(3) and (4) are somewhat tangled, and one needs to choose
convenient variables to disentangle them and to determine the
photon phase space. In spite of the obvious fact that these
constraints concern only the photons, their respective contribution to
the RC affects the acceptance factor given by Eq.~(27). The
physical reason for such a behaviour is that now (in contrast with
previous cases) the 3--momenta of photons and pions do not lie in
the same plane even if the QRE kinematics ($k_1=(1-x)p_1)$ is
applied.

Nevertheless, the QRE approach for the description of collinear
photons allows to simplify the form of the cross section and
to disentangle all the kinematical restrictions. In accordance with
this approach, the starting point for our calculation of the
differential cross section,  suitable for the unrestricted pion phase
space case, is the following
\begin{equation}\label{39}
d\sigma^{^H}_{3F} =
\frac{\sigma(q^2)}{4E^2}\frac{\alpha}{2\pi}P(x,L_0)L^{\gamma}_{\mu\nu}
(xp_1,p_2,k_2)\tilde
g_{\mu\nu}\frac{\alpha}{4\pi^2}\frac{dxd^3k_2}{x\omega_2}\ ,
\end{equation}

$$\frac{d^3k_2}{\omega_2} =2\pi\omega_2d\omega_2dc_2\ , \ \ c_2
=\cos{\theta_2}\ , \ \ x=1-\frac{\omega_1}{E}\ , $$ where
$\theta_2$ is the polar angle of the non--collinear photon. Since
our aim is to derive the differential distribution in the
squared pion invariant mass $q^2$,  it is convenient to use the
relation between $q^2$ and $c_2$ to avoid the integration over $c_2$
on the right--hand side of Eq.~(39). In addition, it is convenient
to introduce the total photon energy $\Omega$ instead of
$\omega_2$
\begin{equation}\label{40}
q^2=4E(E-\Omega) + 2\omega_1\omega_2(1-c_2), \ \
\omega_2=\Omega-\omega_1, \ dc_2\rightarrow\frac{dq^2}{2\omega_1\omega_2}, \ \
d\omega_2=d\Omega\ .  \end{equation}

In this case in the lepton tensor $L^{\gamma}_{\mu\nu}$ the
electron mass can be neglected. Thus, the differential
cross--section has the following form
\begin{equation}\label{41}
\frac{d\sigma^H_{3F}}{dq^2}=\frac{\sigma(q^2)}{4E^2}\Bigl(\frac{\alpha}{2\pi}
\Bigr)^2\Bigl\{P(x,L_0)\Bigl[\frac{2q^4}{xu_1u_2}-2q^2\Bigl(\frac{1}{xu_1}
+\frac{1}{u_2}\Bigr)+\frac{xu_1}{u_2}+\frac{u_2}{xu_1}\Bigr]\frac
{d\omega_1d\Omega}{\omega_1(E-\omega_1)}\Bigr\}\ ,
\end{equation}
$$u_1=-2k_2p_1=-\frac{4E^2\Omega_z}{\omega_1}, \ \
u_2=-2k_2p_2=-\frac{4E}{\omega_1}[\omega_1(\Omega-\omega_1)-E\Omega_z],
\ \ \Omega_z=\Omega-E(1-z)\ . $$

It is useful to rewrite the expression in curly brackets in a form
which is convenient for the integration over $\omega_1$ and
$\Omega$
\begin{multline*}
\Bigl\{-L_0-\frac{zE^2L_0}{(E-\omega_1)^2}+\frac{[2z-(1+z)L_0]E}{E-\omega_1}
+\\
\frac{[2(1+z^2)L_0-4z-(1-z)^2]E^2-(1-z)(\Omega-2\omega_1)E}{\omega_1(\Omega-
\omega_1)-E\Omega_z}\Bigr\}\frac{d\omega_1d\Omega}{E\Omega_z}\ ,
\end{multline*}
 where we neglect terms which do not contain in the denominator the small
quantity $\Omega_z$, which is of
order $\eta E$, as one can see   below from the
expressions for $\Omega_{min}$ and
$\Omega_{max}.$

Our task now is to define the integration region on the
right--hand side of Eq.~(41), which is determined by restrictions
(3) and (4) for the event selection, as well as by the inequalities
\begin{equation}\label{42}
-c_0'< c_2< c_0 \ , \ \ E\Delta <\omega_1 <\Omega-E\Delta, \ \
c_0'=\cos{\theta_0'}, \end{equation} limiting the possible angles
of the non-collinear photon and energies of the collinear one.
Restriction (3) defines the maximum value of $\Omega$, whereas
restriction (4) defines the minimum value of $\omega_1$ at fixed
$\Omega$
\begin{equation}\label{43}
\Omega_{max}=E(1-z)\bigl(1+\frac{\eta}{2}\bigr)\ , \ \
\omega_{min}=\frac{2E\Omega_z}{\Omega-|{\bf K}|c_0}\ , \ \ |{\bf K}|=
\sqrt{\Omega^2-4E\Omega_z}.
\end{equation}

To obtain the minimum value of $\Omega$ we use the first  relation in
(40) at the minimum possible value of $\omega_2=\Delta E$ and $c_2=c_0$
$$\Omega_{min}\Delta(1-c_0)=2E[\Omega_{min}-E(1-z)] $$
which leads to
\begin{equation}\label{44}
\Omega_{min}=E(1-z)\bigl(1+\frac{\Delta(1-c_0)}{2}\bigr)\ .
\end{equation}
From the condition $c_2>-c_0'$ it follows also that
\begin{equation}\label{45}
\omega^-<\omega_1<\omega^+, \ \
\omega^{\pm}=\frac{\Omega}{2}\bigl[1\pm\sqrt{1-\frac{4E\Omega_z}{\Omega^2}
\bigl(1+\frac{1-c_0'}{2}\bigr)}\bigr]\ .
\end{equation}
Finally, inequality $c_2<c_0$ reads: if the values of $\Omega'$s are
such that $$\Omega\leq\Omega_c \ , \ \ \Omega_c =
E(1-z)\Bigl(1+\frac{(1-c_0)(1-z)} {8}\Bigr) $$ then
\begin{equation}\label{46}
\omega_1>\omega_+ \ \ or \ \ \omega_1<\omega_- , \ \
\omega_{\pm}=\frac{\Omega}{2}\bigl[1\pm
\sqrt{1-\frac{8E\Omega_z}{\Omega^2(1-c_0)}}\bigr].
\end{equation}

To derive the integration region we have to combine
consistently all constraints (42)--(46) on $\omega_1$ and $\Omega$, and
such combination leads to
$$\omega_{min}<\omega_1<\omega_- \ \ and \ \
\omega_+<\omega_1< \Omega-\Delta E, \ \
\Omega_{min}<\Omega<\Omega_{\Delta}\ , \ \
\Omega_{\Delta} =E(1-z)(1+\Delta)\ , $$
\begin{equation}\label{47}
\omega_{min}<\omega_1<\omega_- \
\ and \ \ \omega_+<\omega_1< \omega^+, \
\Omega_{\Delta}<\Omega<\Omega_c,
\end{equation}
$$\omega_{min}<\omega_1<\omega^+, \ \
\Omega_c<\Omega<\Omega_{max}. $$ The integration region defined by
the inequalities (47) is shown in Fig. 2.

\begin{figure}[h]
\includegraphics[width=0.7\textwidth]{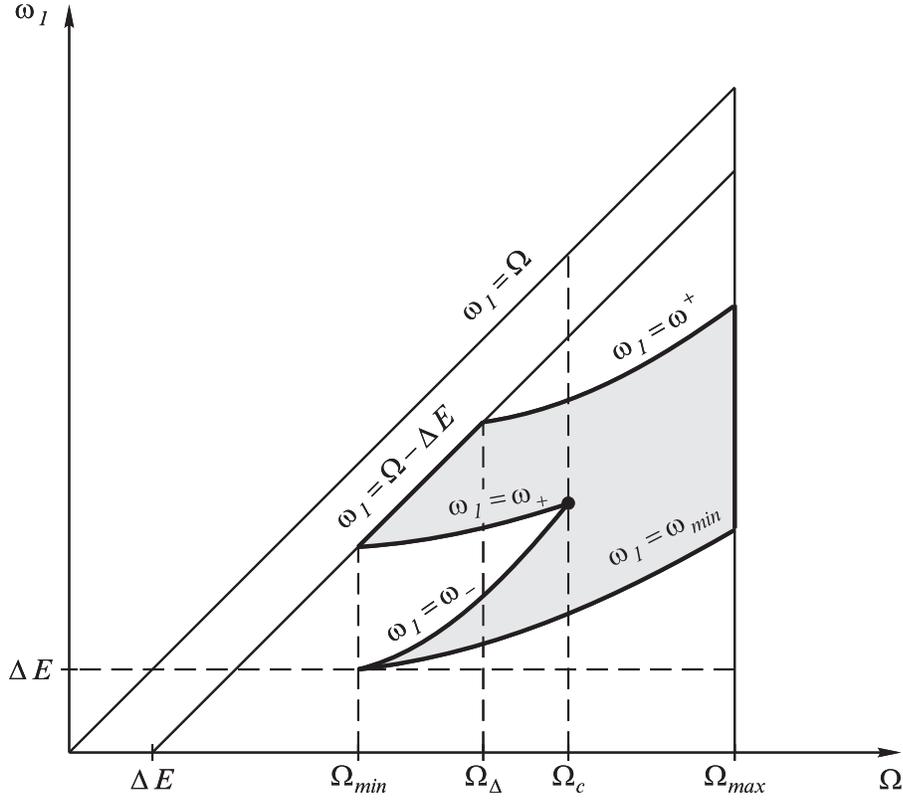}
\caption{The integration region with respect to $\omega_1$ and $\Omega$,
as given by inequalities (47)}\label{fig2}
\end{figure}
The integration with respect to $\omega_1$ and $\Omega$ on the
right--hand side Eq.~(40) over the region (47) can be performed
analytically, and the list of necessary integrals is
$$\int\frac{d\omega_1d\Omega}{E\Omega_z}=(1-z)\bigl(2-\ln\frac{1+\xi}{\xi}
\bigr), \ \ \ \xi=\frac{\eta}{(1-c_0)(1-z)}, $$
$$\int\frac{d\omega_1d\Omega}{(E-\omega_1)\Omega_z}=-\ln{z}\ln\xi+Li_2(1-z)
-Li_2(-\xi z)+Li_2(-\xi)-Li_2\bigl(-\frac{1-z}{z}\bigr), $$
\begin{equation}\label{48}
\int\frac{Ed\omega_1d\Omega}{(E-\omega_1)^2\Omega_z}=
\frac{1}{z}\bigl[
-(1+z)\ln{z}-(1-z)\ln\frac{1+\xi z}{\xi}\bigr],
\end{equation}
$$\int\frac{(\Omega-2\omega_1)d\omega_1d\Omega}{[\omega_1(\Omega-\omega_1)
-E\Omega_z]\Omega_z}=-\frac{\pi^2}{6}-\frac{1}{2}\ln^2\frac{1-c_0}{2}+
\ln\frac{\eta}{2\Delta}\ln\frac{(1-c_0)(1-c_0')}{4}-2Li_2(-\xi),
$$ $$\int\frac{Ed\omega_1d\Omega}{[\omega_1(\Omega-\omega_1)
-E\Omega_z]\Omega_z}=\frac{1}{1-z}\bigl[\frac{\pi^2}{2}+\frac{1}{2}\ln^2
\frac{1-c_0}{2}-
\ln\frac{\eta}{2\Delta}\ln\frac{(1-c_0)(1-c_0')}{4}-\ln^2\xi\bigr]
\ .$$

Using these integrals we can write  the corresponding contribution to the
IES cross-section from the events presently considered,
 as follows
\begin{equation}\label{49}
\frac{d\sigma^{^H}_{3F}}{dq^2}=\frac{\sigma(q^2)}{4E^2}\Bigl(\frac{\alpha}{2\pi}
\Bigr)^2\bigl[2P(z,L_0)G_1+L_0G_2+G_3\bigr]\ ,
\end{equation}
$$G_1=\frac{\pi^2}{2}+\frac{1}{2}\ln^2\frac{1-c_0}{2}-\ln\frac{\eta}{2\Delta}
\ln\frac{(1-c_0)(1-c_0')}{4}-\ln^2\xi \ , $$ $$G_2=(1-z)\bigl[-2+
\ln\frac{(1+\xi)(1+\xi z)}{\xi^2}\bigr]+ $$
$$+(1+z)\bigl[(1+\ln\xi)\ln{z}+Li_2(-\xi
z)+Li_2\bigl(-\frac{1-z}{z}\bigr) -Li_2(-\xi)-Li_2(1-z)\bigr], $$
$$G_3=-(1-z)\bigl(\frac{\pi^2}{3}-\ln^2\xi\bigr)+2Li_2(-\xi)+2z\bigl[
Li_2(1-z)-Li_2(-\xi z)-Li_2\bigl(-\frac{1-z}{z}\bigr)\bigr]. $$

We would like to emphasize that only the part of the
IES cross section, defined by Eq.~(49), has a non--trivial
dependence on the physical parameters $\theta_0$ and $\eta$, which
determine the main requirements for the event selection.

\subsection{Emission of one hard collinear and one hard wide angle
photon with restricted pion phase space}
Consider now the situation with restricted pion phase space. Unfortunately,
the calculations in this case are not so simple and cannot be performed
analytically. Nevertheless, the dependence on the unphysical auxiliary
parameters $\Delta$ and $\theta_0'$, which have to vanish in final result
for total RC, can be extracted.

Our starting point is the following representation for the
differential cross section  corresponding to the QRE approximation
(by analogy with (39))
\begin{equation}\label{50}
d\sigma^{^H}_{3R} =
\frac{12\sigma(q^2)}{4E^2q^2\zeta}\frac{\alpha^2}{8\pi^2}
P(x,L_0)\frac{dx}{x}L^{\gamma}_{\mu\nu}(xp_1,p_2,k_2)(-\tilde p_{-\mu}\tilde
p_{-\nu}) \frac{d^3k_2}{\omega_2}\frac{d\varphi_-}{2\pi}Hdc_-\ ,
\end{equation}
$$H=\frac{|{\bf p}_-|^2}{(2E-\Omega)|{\bf
p}_-|+E_-[\omega_1c_-+(\Omega- \omega_1)c_{2_-}]} \ , \ \ c_{2_-}
= c_2c_-+s_2s_-\cos{(\varphi_2-\varphi_-)}\ , \
s_2=\sin{\theta_2}, $$ where $\varphi_2$ is the azimuthal angle of
the non--collinear photon.

The experience of previous calculations in the case of unrestricted
pion phase space suggests  that, in order  to express the pion
energy via angles and photon energies, we can use $\Omega=E(1-z),$
neglecting only small terms of  order $\eta$ as compared with
unity. Therefore, we have
\begin{equation}\label{51}
E_- =\frac{2Ez(\bar{A}-\bar{K}\bar{B})}{\bar{A}^2
-\bar{B}^2}\ , \ \
\bar{K}=\sqrt{1-\frac{\delta^2}{z^2}(\bar{A}^2-\bar{B}^2)}\ , \ \
\bar{A}=1+z\ , \ \ \bar{B} =(1-x)c_-+(x-z)c_{2-}\ ,
\end{equation}
and with the same accuracy
\begin{equation}\label{52}
H=\frac{2z(\bar{A}\bar{K}-\bar{B})^2}{\bar{K}(\bar{A}^2-\bar{B}^2)^2}\ , \ \
E_+ =E(1+z)-E_-\ .
\end{equation}

In the limiting case when the energy of the non--collinear photon
approaches zero $(x=z),$  $H$ becomes twice the corresponding
value entering, under integral sign, into the expression for the
acceptance factor (27)
\begin{equation}\label{53}
\lim\limits_{x\to z}H = H_s
=\frac{2z[(1+z)K-(1-z)c_-]^2}{K[(1+z)^2-(1-z)^2c_-^2]^2}\ .
\end{equation}

The upper limit of the $c_-$ variation has to be determined from
the 3--momentum conservation, provided that
$\theta_+=\pi-\theta_m,$ that is
\begin{equation}\label{54}
-|{\bf p}_+|c_m+ |{\bf p}_-|c_- +E[1-x+(x-z)c_2] =0\ ,
\end{equation}
where one has to use expressions for $E_-$ and $E_+$ to find
$|{\bf p}_+|$ and $|{\bf p}_-|$ as given by (51) and (52).

Contracting the indices on the right--hand side of Eq.~(50) and
using relations (40) we arrive at the distribution over the
squared di--pion invariant mass
\begin{equation}\label{55}
\frac{d\sigma^{^H}_{3R}}{dq^2}=\frac{12\sigma(q^2)}{4E^2\zeta}
\Bigl(\frac{\alpha}{2\pi}\Bigr)^2P(x,L_0)\frac{d\omega_1d\Omega}{(E-\omega_1)
\omega_1}\frac{TH}{2}\frac{d(\varphi_2 -\varphi_-)}{2\pi}dc_-\ , \
\ x=1-\frac{\omega_1}{E}\ ,
\end{equation}
$$T=-\frac{\delta^2}{z}\Bigl[\frac{(q^2-xu_1)^2+(q^2-u_2)^2}{xu_1u_2}
\Bigr]+\frac{\chi_1(q^2-u_2)}{u_1u_2}
+\frac{\chi_2(q^2-xu_1)}{xu_1u_2}-
\frac{x^2\chi_1^2+\chi_2^2}{xu_1u_2}\ . $$

To advance further, one must integrate first over $c_-$ because
the upper limit $c_{max}$ depends on $c_2$ (as it follows from
(54)), and in its turn $c_2$ is a function of $q^2$ and of the
variables $\omega_1$ and $\Omega.$ In the general case we cannot
integrate analytically, even to write the analytical expression
for $c_{max}$ is a problem. But it is necessary to
 prepare the expressions
  which can then be integrated numerically. Therefore, the
dependence on the unphysical parameters $\Delta$ and $\theta_0'$ has to be
extracted.

For this goal, note first that $\Delta E$ is the minimum possible
energy for the non--collinear photon. Therefore, in order to
extract the $\Delta$--dependence it is sufficient  to investigate
the limit $\omega_2\rightarrow 0.$ In this limiting case $x=z$, and
we can use Eq.~(53) for the term $H$. Moreover, the 3--momentum
conservation (54) becomes the same as in the Born approximation,
and its solution $c_{max}(\omega_2\rightarrow 0)$ coincides with
the expression given in (28).

Next, we select terms in the expression for $T$ that are singular
in this soft limit, because only such terms lead to the
$\Delta$--dependence via $\ln{\Delta}.$ From the list of
integrals (48) and the expression for $u_1, \ u_2$
(see (41)), it is easy to understand that only the terms
containing the product $u_1u_2$ in the denominator can induce
such a dependence.
 Taking
into account also that, in this limiting case,
$\chi_2=q^2-z\chi_1,$ the necessary terms can be written in the
following form
\begin{equation}\label{56}
T_s=\frac{2U16E^2}{u_1u_2}= \frac{2E^3z(1-z)^2U}{\Omega_z[\omega_1(\Omega-
\omega_1)-E\Omega_z]}
\end{equation}
(for the definition of $U$ see Eq.~(27)). To extract the
$\ln{\Delta}$--dependence we apply the standard subtraction
procedure and rewrite the cross section (55) as the sum of its
hard and soft parts
\begin{equation}\label{57}
\frac{d\sigma^{^H}_{3R}}{dq^2}=
\frac{d\sigma^{^{H_h}}_{3R}}{dq^2}+\frac{d\sigma^{^{H_s}}_{3R}}{dq^2}\
,
\end{equation}
where
\begin{equation}\label{58}
\frac{d\sigma^{^{H_s}}_{3R}}{dq^2}
=\frac{12\sigma(q^2)}{4E^2\zeta}
\Bigl(\frac{\alpha}{2\pi}\Bigr)^2P(z,L_0)\frac{d\omega_1d\Omega T_sH_s}{
2E^2z(1-z)}dc_-\ ,
\end{equation}
and the upper limit of $c_-$ variation in Eq.~(58) is defined by
Eq.~(28). The hard part of the cross section on the right--hand
side of Eq.~(57) is not singular at $\omega_2\rightarrow 0$,
whereas the integration of the soft part over the region (47)
induces all $\Delta$--dependence. Using the corresponding formula
in the list of integrals (48) we present the soft part of cross
section (55) in the form
\begin{equation}\label{59}
\frac{d\sigma^{^{H_s}}_{3R}}{dq^2}
=\frac{\sigma(q^2)}{4E^2}
\Bigl(\frac{\alpha}{2\pi}\Bigr)^2P(z,L_0)G_1A(z,c_m)\ ,
\end{equation}
where $G_1$ is defined in Eq.~(49).

It is worth  noting that this soft part absorbs also all the
dependence on the angular auxiliary parameter $\theta_0'.$ That is
the reason why the hard part of the cross section depends on the
physical parameters only and can be computed numerically. The soft
part (58) of the cross section has to be added to the other
contributions into the RC to eliminate the dependence on the
unphysical parameters in the total RC.


\section{Total radiative correction}

\hspace{0.7cm}

The total contribution from radiative corrections  to the Born cross
section in the case of unrestricted
pion phase space is represented by the sum
\begin{equation}\label{60}
\frac{d\sigma^{^{RC}}_{F}}{dq^2}= \frac{d\sigma^{^{S+V}}_{F}}{dq^2}
+\frac{d\sigma^{^{H}}_{1F}}{dq^2} +\frac{d\sigma^{^{H}}_{2F}}{dq^2}
+\frac{d\sigma^{^{H}}_{3F}}{dq^2}\ ,
\end{equation}
whereas for the case of restricted phase space we have
\begin{equation}\label{61}
\frac{d\sigma^{^{RC}}_{R}}{dq^2}= \frac{d\sigma^{^{S+V}}_{R}}{dq^2}+
\frac{d\sigma^{^{H}}_{1R}}{dq^2} +\frac{d\sigma^{^{H}}_{2R}}{dq^2}+
\frac{d\sigma^{^{H_s}}_{3R}}{dq^2} + \frac{d\sigma^{^{H_h}}_{3R}}{dq^2}\ .
\end{equation}

The two auxiliary parameters, the infrared cut--off $\Delta$ and
the collinearity angle $\theta_0',$ enter into the individual
terms on the right--hand sides of Eq.~(60) and (61) in the same
combination
\begin{equation}\label{62}
2P(z,L_0)\ln\Delta\bigl[2(L_s-1)-(L_0'-1)-(L_0-1)+\ln\frac{\theta_0^2
{\theta'}_0^2}{16}\bigr]\ ,
\end{equation}
(here and below we use the expansion of $c_0$ and $c_0'$).
According to the definition of the large logarithms $L_0, \ L_s$ and
$L_0'$ (see Eqs.~(17), (31) and (32)), the expression in square
brackets in Eq. (62) equals to zero, and, therefore, the total RC
depends only on physical parameters and can be written, in the
case of unrestricted phase space, as
\begin{equation}\label{63}
\frac{d\sigma^{^{RC}}_F}{dq^2}=\frac{\sigma(q^2)}{4E^2}\frac{\alpha}{2\pi}
P(z,L_0)\delta^{^{RC}}_F, \ \ \delta^{^{RC}}_F =\frac{\alpha}{2\pi}\frac{
F_0+L_0F_1+F_2}{P(z,L_0)}\ ,
\end{equation}
$$F_0=
\frac{1}{2}L_0^2P_{2\theta}(z)+P(z,L_0)\bigl[L_s\bigl(
\frac{3}{2}+2\ln\frac{\eta}{2}\bigr)+\ln\frac{4}{\theta_0^2}
\bigl(\frac{3}{2}+2\ln\frac{\eta}{z\theta_0}\bigr)
-2\ln^2\xi-2\ln\frac{\eta}{2}+3\ln{z}+\frac{5\pi^2}{3}-\frac{9}{2}\bigr]
 \ , $$
$$F_1=\frac{3-8z+z^2}{2(1-z)}-\frac{2(1+z)^2}{1-z}\ln(1-z)+\bigl[\frac{4z}
{1-z}+(1+z)(1+\ln\xi)\bigr]\ln{z} +\frac{1}{2}(1+z)\ln^2z
+\frac{1+z^2}{1-z} \bigl(\frac{\pi^2}{3}-$$ $$-2Li_2(z)\bigr)+
(1-z)\ln\frac{(1+\xi)(1+\xi z)}{\xi^2}+(1+z)\bigl[ Li_2(-\xi
z)+Li_2\bigl(-\frac{1-z}{z}\bigr)-Li_2(-\xi)-Li_2(1-z)\bigr], $$
$$F_2=(1-z)\ln^2\xi-\frac{3+4z}{3(1-z)}+\frac{4z}{1-z}\ln(1-z)+\bigl[-2z
\ln\xi+\frac{3-18z+7z^2}{3(1-z)^2}\bigr]\ln{z}+2z\ln^2(1-z)+$$
$$+\frac{3-12z+30z^2-36z^3+7z^4}{6(1-z)^3}\ln^2z+\frac{\pi^2}{6}
\bigl(4+\frac{14z}{3}-\frac{5}{1-z}\bigr)+Li_2(z)\bigl(4z-6
+\frac{5}{1-z}\bigr)+$$
$$+\bigl(1-\frac{z}{3}\bigr)Li_2(1-z)+2Li_2(-\xi)-2z\bigl[Li_2(-\xi
z) +Li_2\bigl(-\frac{1-z}{z}\bigr)\bigr]+J\ . $$

As  mentioned in the Introduction, the IES Born cross section and
its RC factorize into the low energy pion pair production
cross--section, $\sigma(q^2),$  and a term of a  pure
electrodynamical origin. This latter term depends on the measured
pion invariant mass $q^2$ and on the physical parameters $\eta$
and $\theta_0$ which define the rules for the IES.

\begin{figure}[h]
\includegraphics[width=0.7\textwidth]{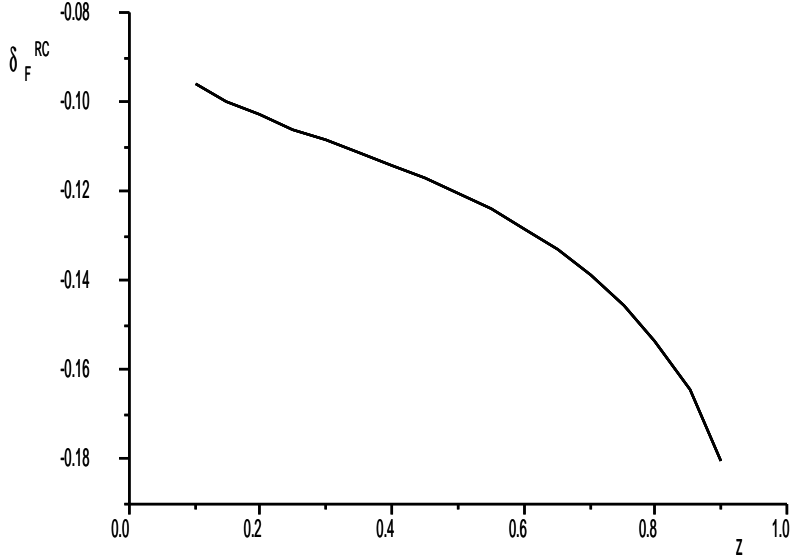}
\caption{The full first order radiative correction to the Born
cross section (21) for the case of unrestricted pion phase space, as
defined by Eq.~(63)}\label{fig3}
\end{figure}

The $z$--dependence of  the total radiative
correction to the Born cross section (21),$\delta^{^{RC}}_F$,
is shown in Fig. 3.
Note that the contribution of the non--logarithmic terms to
$\delta^{^{RC}}_F$
 equals parametrically to $\alpha/(2\pi L_0)$, which  is  of order
$10^{-4}$, as the  relative contribution of terms proportional to
${\bf P}_{\Phi}^2/4E^2$ and $\theta_0^2/6$ in the Born cross
section (see Eq.~(20)). That is why the exact calculation of the Born IES
cross-section should be complemented with the radiative corrections
calculated with the
inclusion of non--logarithmic contributions.

Similarly to (63), we can write the complete radiatively corrected cross-section
for the case of restricted
pion phase space in the form
\begin{equation}\label{64}
\frac{d\sigma^{^{RC}}_R}{dq^2}=\frac{\sigma(q^2)}{4E^2}\frac{\alpha}{2\pi}
P(z,L_0)A(z,c_m)\delta^{^{RC}}_R + \frac{d\sigma^{^{H_h}}_{3R}}{dq^2}\ ,
\end{equation}
where
\begin{equation}\label{65}
\delta^{^{RC}}_R
=\frac{\alpha}{2\pi}\frac{F_0+L_0F_{1R}+F_{2R}}{P(z,L_0)} \ \ ,
F_{1R}=F_{1}-G_2\ , \ \ F_{2R}=F_{2}-G_3 \ .
\end{equation}

Because the factor $\sigma(q^2)/4E^2$ enters into the last term at
the right--hand side of Eq.~(64) also, the total RC in this case
has a factorized form as well.

There exists one more contribution   caused by two hard photon
emission when   neither photon is emitted within
 the narrow cone along the
the electron beam direction but, nevertheless, the collinear
condition (4) is satisfied. This contribution cannot be calculated
by the QER approach and has to be evaluated by other methods. In
particular, the double bremsstrahlung lepton current tensor can be
taken in limit $m\rightarrow 0.$ To our understanding,
due to the strong
constraint (3) on the event selection, the corresponding
contribution is small enough and does not  affect the IES cross section on
the one per cent level. Nevertheless, the theoretical evaluation
of this contribution should be done and we hope to compute it
elsewhere.

\section{Pair production contribution into the IES cross section}

\hspace{0.7cm}

The above considerations for the photonic radiative corrections
to the IES cross
section are appropriate if $e^+e^-\pi^+\pi^-$ final states are
excluded from the analysis. If not, there is an additional
contribution caused by hard initial--state radiation with $e^+e^-$ pair
production \cite{AKMT}. The main part of this contribution arises
due to collinear kinematics. In the framework of the NLO
approximation, where
 only logarithmically enhanced  terms are kept, the corresponding cross
section can be written as
\begin{equation}\label{66}
\frac{d\sigma^{e^+e^-(c)}_F}{dq^2}=\frac{\sigma(q^2)}{4E^2}\Bigl(\frac{\alpha}
{2\pi}\Bigr)^2\bigl[P_1(z)L_0^2+P_2(z)L_0\bigr]\ ,
\ \ \frac{d\sigma^{e^+e^-(c)}_R}{dq^2}=
\frac{d\sigma^{e^+e^-(c)}_F}{dq^2}A(z,c_m)\ ,
\end{equation}
where the functions $P_1(z)$ and $P_2(z)$ can be extracted from
the corresponding results for small--angle Bhabha scattering
cross--section, given in Ref. \cite{Bhabha}. We present them here
for completeness, i.e.
$$P_1(z)=\frac{1+z^2}{3(1-z)}+\frac{(1-z)(4+7z+4z^2)}{6z}+(1+z)\ln{z}\
, $$ $$P_2(z)=-\frac{107}{9}+\frac{136}{9}z-\frac{2}{3}z^2
-\frac{4}{3z}
-\frac{20}{9(1-z)}+\frac{2}{3}\Bigl(-4z^2+5z+1+\frac{4}{z(1-z)}\Bigr)\ln(1-z)$$
$$+\frac{1}{3}\Bigl(8z^2+5z-7-\frac{13}{1-z}\Bigr)\ln{z}-\frac{2}{1-z}\ln^2z
+4(1+z)\ln{z}\ln(1-z)-\frac{2(3z^2-1)}{1-z}Li_2(1-z)\ .$$

Within the NLO accuracy, one has to compute also the contribution
caused by the semicollinear kinematics of the $e^+e^-$ pair
production, when the final state electron belongs to the narrow cone
along the electron beam direction while the positron does not. The
corresponding part of the leptonic tensor was derived in
\cite{KMS}, and has the  form $$L_{\mu\nu}^{e+e-}=
-\frac{\alpha^2}{4\pi^3}\frac{d^3k_+
d\,x(1+x^2)}{\varepsilon_+(1-x)^2sv_1}L_0
\Bigl\{\Bigl[\frac{s^2+v_1^2}{2}+\frac{q^2v_2}{(1-x)^2}\Bigr]\tilde
g_{\mu\nu}+ \frac{2q^2}{(1-x)^2}(\tilde p_{2\mu}\tilde p_{2\nu} +
\tilde k_{+\mu}\tilde k_{+\nu})\Bigr\} \ , $$ where $k_+ \
(\varepsilon_+)$ is the 4 -- momentum (energy) of the
non--collinear positron, $x$ is the energy fraction of the
collinear electron and $v_{1,2}=-2p_{1,2}k_+.$

For unrestricted pion phase space, the
differential cross section can be written as follows
\begin{equation}\label{67}
\frac{d\sigma^{e+e-(s)}_F}{dq^2} =
\frac{\sigma(q^2)}{4E^2}\Bigl(\frac{\alpha}{2\pi}\Bigr)^2L_0\frac{d\Omega
dx}{\Omega_z}\frac{(1+x^2)[(1-x-z)^2+z^2]}{(1-x)^4} \ ,
\end{equation}
Neglecting terms of  order $m/E,$ we can integrate over the region
shown in  Fig.~2 with the substitution $\omega/E\rightarrow x.$ In addition
to the list of integrals (48), it is necessary to compute the following
ones $$\int\frac{dx\,d\Omega}{(1-x)^3\Omega_z} =
\frac{(1-z)^2\xi}{2z(1+z\xi)} +\frac{1-z^2}{2z^2}
\Bigl(1+\ln\frac{1+z\xi}{\xi}\Bigr) -\frac{1+z^2}{2z^2}\ln{z} \
,$$
\begin{equation}\label{68}
\int\frac{dx\,d\Omega}{(1-x)^4\Omega_z}
=\frac{(1-z)(3+z)}{3z^3}+\frac{(1-z)^3}{6z^3(1+z\xi)^2}-\frac{(1-z)^2(2+z)}{3z^3(1+z\xi)}-
\end{equation}
$$-\frac{1+z^3}{3z^3}\ln{z}
-\frac{1-z^3}{3z^3}\ln\frac{1+z\xi}{\xi}\ . $$
\begin{figure}[t!]
\includegraphics[width=0.7\textwidth]{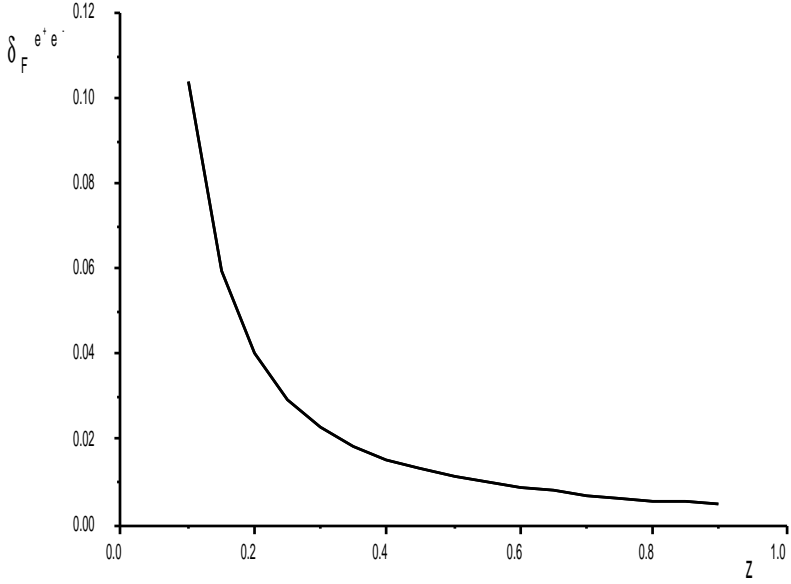}
\caption{The radiative correction to the IES Born cross section
caused by $e^+e^-$--pair production}\label{fig4}
\end{figure}

Using integrals (48), (68) and the definition (67), the
contribution of the semicollinear kinematics of pair production
into the IES cross section can be written in the form
\begin{equation}\label{69}
\frac{d\sigma^{e+e-(s)}_F}{dq^2} =
\frac{\sigma(q^2)}{4E^2}\Bigl(\frac{\alpha}{2\pi}\Bigr)^2L_0
S(z,\xi)\ , \end{equation} $$ S(z,\xi)) = (1-z)\Bigl\{
\ln\frac{\xi}{1+\xi}-\frac{2}{3}\Bigl[1+\frac{(1-z)^2\xi}{(1+z\xi)^2}
+ \frac{2(1+z+z^2)}{z}\ln\frac{1+z\xi}{\xi} \Bigr]\Bigr\} + $$
$$2(1+z)\Bigl[\Bigl(\ln{\xi} -\frac{2}{3z}(1-z+z^2)\Bigr)\ln{z}
+Li_2(-z\xi)+Li_2\bigl(-\frac{1-z}{z}\bigr)-Li_2(1-z)-Li_2(-\xi)
\Bigr] \ . $$

The contribution of $e^+e^-$ pair production into the IES cross section
for the case of unrestricted pion phase space now reads
\begin{equation}\label{70}
\frac{d\sigma^{^{e+e-}}_F}{dq^2}=\frac{\sigma(q^2)}{4E^2}\frac{\alpha}{2\pi}
P(z,L_0)\delta^{^{e+e-}}_F, \ \ \delta^{^{e+e-}}_F
=\frac{\alpha}{2\pi}\frac{ P_1(z)L_0^2+
(P_2(z)+S(z,\xi))L_0}{P(z,L_0)}\ .
\end{equation}
with the function $\delta^{^{e+e-}}_F$  shown in Figure 4.
In the case of restricted pion phase space, the differential distribution over
the pion squared
 invariant mass can be written, in analogy with Eq.~(55), in the
following form
\begin{equation}\label{71}
\frac{d\sigma^{e^+e^-(s)}_R}{dq^2}=\frac{12\sigma(q^2)}{4E^2\zeta}\Bigl(\frac
{\alpha}{2\pi}\Bigr)^2L_0\frac{(1+x^2)d\,xd\Omega}{(1-x)^4\Omega_z}
\frac{T_pH_p}{2}dc_-\frac{d(\tilde{\varphi}_+-\varphi_-)}{2\pi}\ ,
\end{equation}
where the integration region in  $x$ and $\Omega$ is the same as
in (67). Here, we use the following notation
$$T_p=-\frac{\delta^2}{z^2}[z^2+(1-x-z)^2]+x\tilde{\chi}_2+(1-x-z)\tilde{\chi}
_+ +\tilde{\chi}_2^2+\tilde{\chi}_+^2\ , \ \
\tilde{\chi_2}=\frac{\chi_2} {4E^2}\ , \ \ \tilde{\chi}_+ =
\frac{2k_+p_-}{4E^2}\ , $$ $$H_p=\frac{2z(\widetilde A\widetilde
K-\widetilde B)^2}{\widetilde K( \widetilde A^2-\widetilde
B^2)^2}\ , \ \ \widetilde A =A, \ \ \widetilde B= xc_-
=(1-x-z)c_{+-}\ , $$ $$c_{+-}=\tilde{c}_+c_-
+\tilde{s}_+s_-\cos{(\tilde\varphi_+ -\varphi_-)} \ , \ \
\tilde{c}_+=\cos{\tilde{\theta}_+}\ , \ \ \tilde{s}_+ =
\sin{\tilde {\theta}_+}\ , $$ and $\tilde{\theta}_+, \
\tilde{\varphi}_+ $ are polar and azimuthal angles of the final
non-colinear positron.

For the further integration on the right--hand side of
Eq.~(71) with respect to the angular pion phase space, one must
determine the upper limit of $c_-$ variation. It depends on $q^2,
\ x, \ \Omega$ and $\cos{(\tilde{\varphi}_+ -\varphi_-)}$ and can
be obtained as a solution of the equation
\begin{equation}\label{72}
-|{\bf p}_+|c_m +|{\bf p}_-|c_-+E(x+(1-x-z)\tilde{c_+}) = 0\ ,
\end{equation}
taking into account that
$$E_-=\frac{2Ez(\widetilde A -\widetilde K\widetilde B)}{\widetilde A^2
-\widetilde B^2}\ , \ \ E_+=E(1+z)-E_-\ . $$

The total contribution of $e^+e^-$--pair production into IES cross section
for the case of restricted pion phase space
\begin{equation}
\frac{d\sigma^{e+e-}_R}{dq^2} =\frac{d\sigma^{e+e-(c)}_R}{dq^2}+
\frac{d\sigma^{e+e-(s)}_R}{dq^2}
\end{equation}
has no singularity and can be calculated numerically.

We have considered in the above the contribution of kinematic
regions where at least one collinear photon or an electron--positron
pair is radiated by the initial electron. DA$\Phi$NE conditions
allow to select and detect also the same events when the collinear
particles are emitted by the initial positron. Therefore, all the
cross sections derived above have to be doubled.

\section{Conclusion}

\hspace{0.7cm}

The success of  precision studies of the  hadronic cross section in
electron--positron annihilation through the measurement of
radiative events \cite{MEMO,Raz1,D} relies on the matching level of
reliability of the theoretical expectations. The principal problem
is the analysis of radiative corrections corresponding to
realistic conditions for event selection.

In previous work \cite{mb1} we discussed briefly an inclusive approach
to the measurement of the hadronic cross section at DA$\Phi$NE in
the region below $1\ GeV$ by the radiative return method, for the
case in which the radiated photon remains untagged. This approach requires
an exact
knowledge of the final hadronic state and a precise determination
of its invariant mass. Some additional constraints have to be
imposed to make the detection of the ISR photon redundant and to
avoid any uncertanties in the interpretation of the selected
events. These additional constraints imply also the precise
measurement of the pion 3--momenta. The KLOE detector at DA$\Phi$NE offers
a very promising possibility to realize such an inclusive approach to
the scanning of the hadronic cross section by the ISR events.

In this paper we compute the corresponding ISR Born cross section
and the radiative corrections to it in the framework of the QRE
approximation. It is shown that this approximation is quite
appropriate, even at the Born level, and provides high accuracy for
the IES cross section. The cases of unrestricted and restricted pion
phase space are considered. In the first case the photonic
contribution to the RC is calculated analytically with the NNLO
accuracy and the contribution caused by $e^+e^-$--pair production
within the NLO. The photonic RC is large and negative in a wide
range of  pion invariant masses. The physical reason for
such  behaviour of the photonic RC is very
transparent: the phase space of additional real photon is restricted
considerably by the constraints (3) and (4), and the respective
positive contribution cannot compensate the negative contribution
due to the virtual correction. The large absolute value of the
first order photonic correction indicates unambiguously that the
second order RC has to be evaluated. Moreover, the increase of the
soft part of the RC (when $z$ grows and approaches  unity) requires
summation of the leading RC to all orders.

If the entire phase space for photons and $e^+e^-$--pairs is
allowed, this problem is solved by the ordinary Drell--Yan--like
representation in electrodynamics \cite{KF,BvN} with the exponential
form of the electron structure functions \cite{JSW,GL}. For the
tagged collinear photon events without any constraints on the
phase space of additional particles, the corresponding
representation was derived in Ref. \cite{AKMT}, but the case
considered here  requires special investigation because of two
non--trivial constraints (3) and (4) on the event selection. We
hope to consider this problem elsewhere.

The RC caused by $e^+e^-$--pair production is positive and small,
as compared with the absolute value of the  photonic correction. Only
in the region near  threshold (small $z$), where the cross
section is very small, it approaches approximately the same value.
So, we conclude that the RC due to pair production  described by
Eq.~(70) is adequate, and it must be taken into account to
guarantee the one per cent accuracy.

In the case of restricted pion phase space, we derived the analytical
form of the acceptance factor for the Born cross section and the
part of RC which includes contributions due to vitrual and real
collinear photons and $e^+e^-$ pair. Even the choice of the
detected pion angles between $20^o$ and $160^o$ gives a value of
the acceptance factor  very close to unity, providing good
statistics  at all values of the hadron invariant mass.
Concerning the contribution into the IES cross section caused by
the semicollinear kinematics for the double photon emission and
pair production, the respective acceptance factor cannot be
calculated analytically. In this case formulae suitable for
numerical calculation are given.

One of the advantages of the IES approach discussed in this paper
 is  a considerable decrease of the FSR background. The
 corresponding contribution into the IES cross section is
 suppressed by factors of order $\theta_0^2$ due to the collinear
 restriction (4) on event selection, therefore we can ignore any RC
 to the FSR events and evaluate this background only at the Born
 level. The same is valid for the contribution caused by the
 ISR--FSR interference. Note that, for the tagged photon setup, the
 corresponding background is quite large and, in order
to obtain the  one per cent
 accuracy \cite{HGJ},  one needs evaluate
the radiative corrections to it. If events with $e^+e^-\pi^+\pi^-$ final state cannot be
 excluded by the experimental selection,
 the background due to the double
 photon mechanism of $\pi^+\pi^-$--pair production has to be evaluated as
well.
 To our understanding, this background contains the same
 suppression factor $\theta_0^2$. In addition,  restriction (3)
 selects very specific kinematics and does not allow to reach the
 region where the  virtualities  of both intermediate photons are
 small, and the corresponding cross section is the largest. In fact,
 due to this restriction, at least, one of the photons becomes far
off-shell (with virtuality of the order $-q^2$). Thus, we expect
that double photon mechanism contributes at the level of the RC to
the FRS and cannot  affect the IES cross section at the one per
cent level.

Authors thank G. Venanzoni for fruitful discussion. V.A.K. is
grateful to The Leverhulme Trust for a Fellowship. N.P.M. thanks
INFN and Parma University for the hospitality.
 G. P., acknowledges support from EEC Contract TMR98-0169.

\end{document}